\documentclass{article}

\pdfoutput=1

\usepackage{amsmath}
\usepackage[hidelinks]{hyperref}
\usepackage[super]{nth}
\usepackage[nottoc,numbib]{tocbibind}
\usepackage{cite}
\usepackage{siunitx}
\usepackage[labelfont=bf]{caption}
\usepackage{float}
\usepackage[margin=1in]{geometry}
\usepackage{multirow}
\usepackage{rotating}
\usepackage{graphicx}
\usepackage{authblk}
\usepackage{parskip}
\usepackage{multirow}
\usepackage{colortbl}

\newcommand{\email}[1]{\href{mailto:#1}{#1}}

\newcommand{\beginsupplement}{%
        \setcounter{table}{0}
        \renewcommand{\thetable}{S\arabic{table}}%
        \setcounter{figure}{0}
        \renewcommand{\thefigure}{S\arabic{figure}}%
     }

\title{Puzzle Imaging: Using Large-scale Dimensionality Reduction Algorithms for Localization }
\author[1*]{Joshua~I.~Glaser}
\author[2]{Bradley~M.~Zamft}
\author[2]{George~M.~Church}
\author[1,3,4]{Konrad~P.~Kording}
\affil[1]{Department of Physical Medicine and Rehabilitation, Northwestern University and Rehabilitation Institute of Chicago, Chicago, Illinois, USA}
\affil[2]{Department of Genetics, Harvard Medical School, Boston, Massachusetts, USA}
\affil[3]{Department of Physiology, Northwestern University, Chicago, Illinois, USA}
\affil[4]{Department of Applied Mathematics, Northwestern University, Chicago, Illinois, USA}
\affil[*]{\textnormal{Direct correspondence to \email{j-glaser@u.northwestern.edu}}}

\date{}

\begin{document}

\maketitle

\begin{abstract}
Current high-resolution imaging techniques require an intact sample that preserves spatial relationships. We here present a novel approach, ``puzzle imaging," that allows imaging a spatially scrambled sample. This technique takes many spatially disordered samples, and then pieces them back together using local properties embedded within the sample. We show that puzzle imaging can efficiently produce high-resolution images using dimensionality reduction algorithms. We demonstrate the theoretical capabilities of puzzle imaging in three biological scenarios, showing that (1) relatively precise 3-dimensional brain imaging is possible; (2) the physical structure of a neural network can often be recovered based only on the neural connectivity matrix; and (3) a chemical map could be reproduced using bacteria with chemosensitive DNA and conjugative transfer. The ability to reconstruct scrambled images promises to enable imaging based on DNA sequencing of homogenized tissue samples.
\end{abstract}

\tableofcontents

\section{Introduction}
Many biological assays require the loss/destruction of spatial information of samples, making it difficult to create a high-resolution image of cellular properties. As a prime example, determining genetic properties of a biological sample usually requires breaking apart that sample for DNA sequencing (but see~\cite{lee14}). This limits the resolution of an image of genetic content to the precision of tissue sectioning prior to sequencing. Along with determining gene expression, researchers are attempting to use genetic information to determine neural connectivity~\cite{zador12}, neural activity~\cite{kording11}, other cellular properties~\cite{farzadfard14}, and chemical concentrations~\cite{moser13}. Being able to image these types of properties at high resolution and large scale could therefore lead to expanded measurement and recording applications. This would be possible if we could recover spatial information post hoc.

\begin{figure}[H]
\centering
\includegraphics[scale=1]{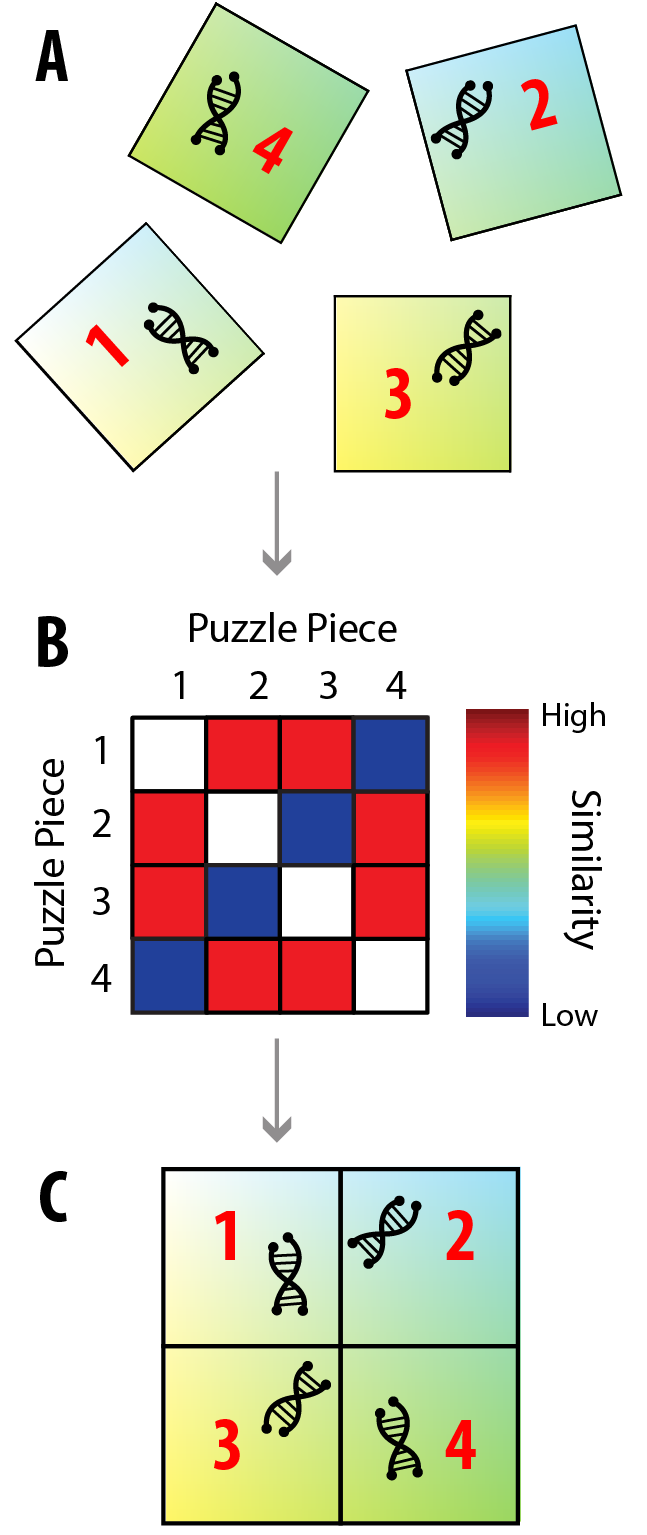}
\caption[Puzzle Imaging]{\textbf{Puzzle Imaging.} There are many properties, such as genetic information, that are easier to determine when the original spatial information about the sample is lost. However, it may be possible to still image these properties using relative spatial information. \textbf{(A)} As an example, let us say that each piece of genetic information is attached to a puzzle piece. While the puzzle pieces don't provide absolute spatial information, they provide relative spatial information: we know that nearby pieces should have similar colors, so we can use color similarity to determine how close puzzle pieces should be to one another. \textbf{(B)} We can make a similarity matrix of the puzzle pieces, which states how similar the puzzle pieces' colors are to each other, and thus how close the pieces should be to one another. \textbf{(C)} Through dimensionality reduction techniques, this similarity matrix can be used to map each puzzle piece to its correct relative location. } \label{fig:1}
\end{figure}

In order to recover a sample's spatial information, information about its relative spatial location could be embedded and utilized. For example, imagine that each piece of genetic information was attached to a puzzle piece (the embedded relative spatial information; \autoref{fig:1}A). While the puzzle pieces by themselves don't provide spatial information, fitting the pieces together would lead to a spatially correct image of the genetic information. Thus, the use of relative spatial information (how the puzzle pieces' locations relate to one another) could allow for higher-resolution imaging.

Using relative spatial information to reconstruct an image can be thought of as a dimensionality reduction problem. If there are $N$ puzzle pieces, one can construct an $N \times N$ similarity matrix $\mathbf{S}$, where $\mathbf{S}_{ij}$ determines how close puzzle piece $i$ and $j$ are (higher similarity means shorter distance; \autoref{fig:1}B). The goal is to map this high dimensional similarity matrix to accurate 2- or 3-dimensional locations of each piece (\autoref{fig:1}C). Importantly, there is a whole class of dimensionality reduction methods that aims to preserve high dimensional distances in the reduced dimension (e.g.~\cite{kruskal64,tenenbaum00,coifman06}). These types of techniques would allow a ``piecing of the puzzle back together."

Here, we propose ``puzzle imaging," and develop two dimensionality reduction algorithms that would allow large-scale puzzle imaging. We describe three concrete examples in which puzzle imaging would be beneficial: (1) ``Neural Voxel Puzzling," in which a relatively high-resolution 3-dimensional brain map is reproduced by giving DNA barcodes to neurons; (2) ``Neural Connectomics Puzzling," in which neural connections are used to recover neural locations; and (3) ``Chemical Puzzling," in which a chemical map could be reproduced using bacteria with chemosensitive DNA and conjugative transfer. Each of these examples leverage the faster-than-Moore's law advances in both the cost and speed of genetic sequencing~\cite{carr09}, and therefore are likely to become more relevant as this ``genetic revolution" proceeds. In each example, we use our algorithms on simulated data, and provide a preliminary demonstration of the capabilities of puzzle imaging. 

\section{Results}
\label{sec:results}

\subsection{Neural Voxel Puzzling}
\subsubsection{Overview}
The purpose of neural voxel puzzling is to create a 3-dimensional image of the brain at high resolution. This image could provide useful neuroanatomical information by itself, or could be used in conjunction with other technologies to determine the locations of genetically encoded neural properties in the brain~\cite{marblestone14}.

The first step in voxel puzzling is to label each neuron in the brain with a unique DNA or RNA barcode throughout its entire length (\autoref{fig:2}A). Ongoing research aims to tackle this challenge~\cite{zador12,peikon14,gerlach13}. Recently, researchers have succeeded in having bacteria generate a large diversity of barcodes \textit{in vivo}~\cite{peikon14}. Next, the brain is shattered into many voxels (\autoref{fig:2}B; note that voxels are only squares for simplification), and the DNA in each voxel is sequenced, yielding a record of which neurons were in which voxels. This provides us with relative spatial information about voxel placement: voxels that share more neurons will likely be closer to each other. We can use this relative spatial information to puzzle the voxels into their correct locations.

\subsubsection{Dimensionality Reduction} 

\begin{figure}[H]
\centering
\includegraphics[scale=1]{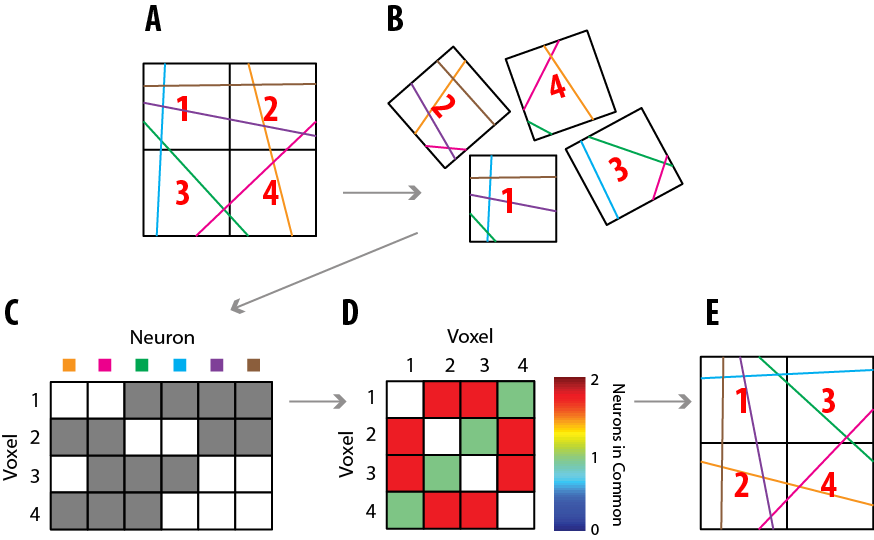}
\caption[Neural Voxel Puzzling Overview]{\textbf{Neural Voxel Puzzling Overview.} \textbf{(A)} An example of 6 ``neurons" (lines) going through 4 voxels. Each neuron has a unique DNA barcode (here color). \textbf{(B)} These voxels are broken apart. \textbf{(C)} A coincidence matrix, $\mathbf{X}$, is constructed describing which neurons are in which voxels. Gray signifies that a neuron is in a particular voxel. \textbf{(D)} A similarity matrix is constructed describing how many neurons a pair of voxels has in common. This matrix can be calculated as $\mathbf{XX}^T$. \textbf{(E)} The voxels are puzzled back together. The reconstruction may be rotated or flipped, as shown here. } \label{fig:2}
\end{figure}

More formally, puzzling together the brain is a dimensionality reduction problem, with each voxel represented by an $N$-dimensional object, where $N$ is the number of neurons.  Dimension $k$ corresponds to neuron $k$ being in that voxel (\autoref{fig:2}C).  These $N$-dimensional voxels must be mapped to their correct 3-dimensional coordinates (or 2 dimensions in Figure 2's simplified example). This can be done because voxels that are closer in 3-dimensional space will have more neurons in common, and thus will be more similar in $N$-dimensional space. With the knowledge of the similarity between all pairs of voxels, we can create a similarity matrix (\autoref{fig:2}D) and then puzzle the voxels back together (\autoref{fig:2}E). Without any additional information, it is impossible to know exact locations; only the relative locations of voxels can be known. Thus, the output could be a flipped or rotated version of the input (\autoref{fig:2}E). That being said, additional information could be used to correctly orient the reconstruction. For example, the overall shape could be matched for a sample that is asymmetric (like the brain), or a few samples with known locations could act as landmarks.

In order to convert the knowledge of which neurons are in which voxels (\autoref{fig:2}C) to a reconstructed puzzle (\autoref{fig:2}E), a dimensionality reduction algorithm is needed. We demonstrate the performance of two algorithms that are promising for large-scale problems. The first is a variant of Diffusion Maps~\cite{coifman06}, which uses a sparse similarity (affinity) matrix instead of the standard calculation. We will refer to this method as Sparse Diffusion Maps, or SDM. The second is a variant of Landmark Isomap~\cite{desilva04,silva02}, which is faster for unweighted graphs (binary similarity matrices). We will refer to this method as Unweighted Landmark Isomap, or ULI. In the below simulations, we use 10 landmark points. For the full algorithms, see \textit{\nameref{sec:methods}}.

\subsubsection{Performance} 

We tested the ability of both dimensionality reduction algorithms to determine the locations of 8000 simulated voxels of varying dimensions. Voxels were not confined to be cubes; they could be any shape. In our simplistic simulations, our ``neurons" were long rods with cross-sectional areas of $\SI{1}{\micro\meter}^2$ (about the size of an axon~\cite{perge12}) and were assumed to fully go through each voxel they entered. We set the total number of neurons in our simulations so that they would fill voxels of the chosen size. Neurons were oriented within the volume at randomly determined angles. See \textit{\nameref{sec:methods}} for simulation details.

In our simulations, we used two metrics to determine the quality of the reconstruction. The first metric was the mean error in distance across all voxels. Because the final reconstruction will be a scaled, rotated, and reflected (flipped) version of the initial voxel locations, we first transformed (scaled, rotated, and reflected) the final reconstruction to match the original voxel locations. We used the transformation that minimized the mean error. 

Another metric we used is the correlation coefficient (R value) of the plot of reconstructed distances between all points against their true distances. This metric determines how well relative distances between points are preserved, and does not depend on any transformations following the reconstruction. A perfect reconstruction (ignoring differences in scaling, rotation, and reflection) would have a linear relationship between the true and reconstructed distances, and thus an R value of 1. 

\begin{figure}[H]
\centering
\includegraphics[scale=.82]{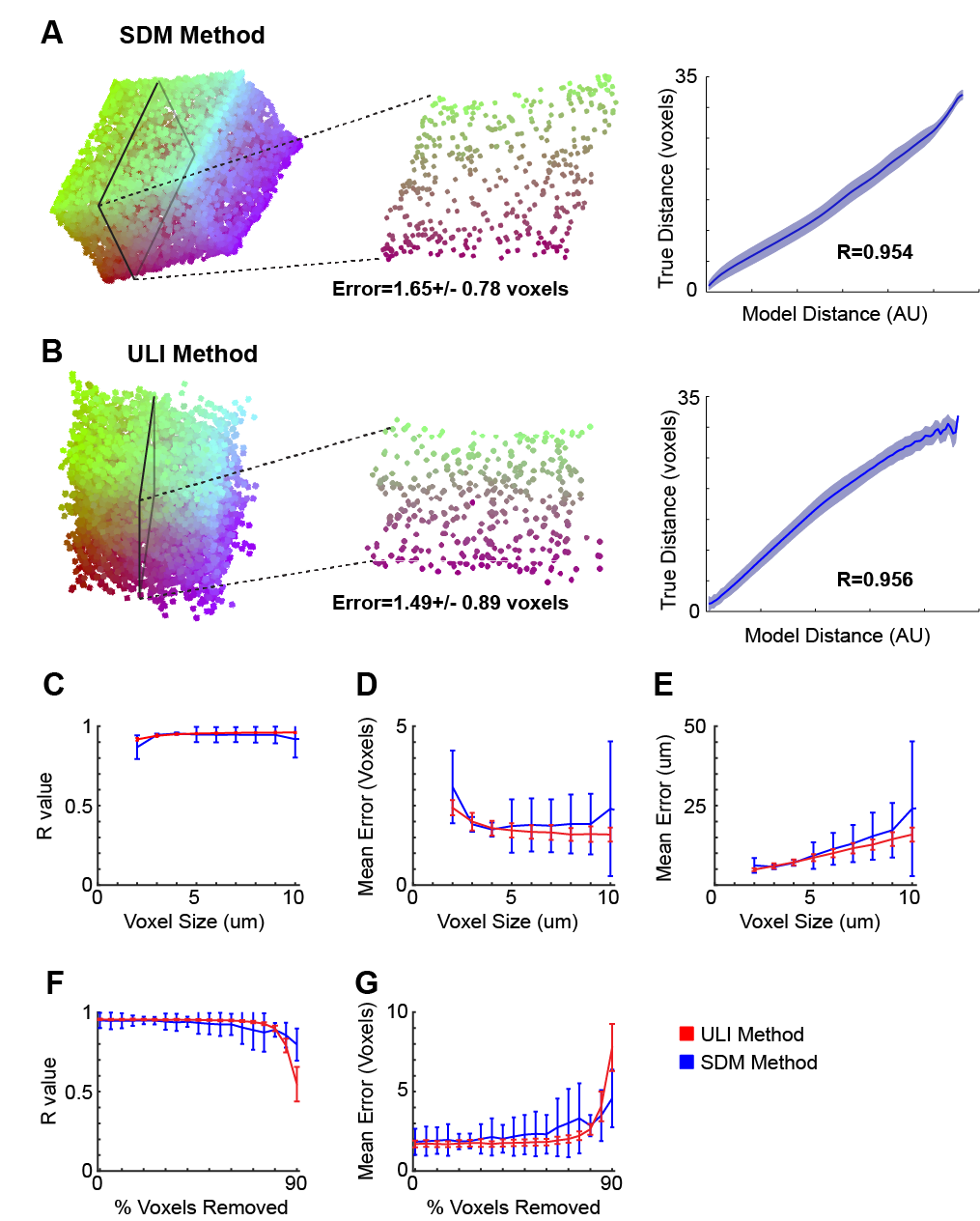}
\caption[Neural Voxel Puzzling Performance]{\textbf{Neural Voxel Puzzling Performance.} \textbf{(A)} On the \textbf{left}, an example reconstruction of voxel locations using the SDM method. Colors are based on initial locations: those with a larger initial $x$ location are redder, while those with a larger initial $y$ location are bluer. In the \textbf{middle}, a 2-dimensional slice through reconstructed volume. The distance errors are calculated following scaling and rotating the reconstructed volume to match the original volume. On the \textbf{right}, one metric for the accuracy of reconstruction is shown by plotting the reconstructed distances between all points against their true distances for the reconstruction in this panel. The mean plus/minus the standard deviation (shaded) is shown. A perfect reconstruction would be a straight line, corresponding to an R value of 1. \textbf{(B)} Same as panel A, except an example reconstruction using the ULI method. \textbf{(C)}  R values for simulations using the SDM method (blue) and ULI method (red), as a function of the voxel size. While voxels were not confined to be cubes, for ease of understanding, we report voxel sizes as the edge length of the cube corresponding to the average voxel size. Error bars represent the standard deviation across simulations in each panel. \textbf{(D)}  Mean distance errors in voxels for both methods as a function of the voxel size. \textbf{(E)}  Mean distance errors in microns for both methods as a function of the voxel size. \textbf{(F)}  Voxels are removed to represent voxels that do not contain location information (such as voxels that contain a single cell body). R values for simulations using both methods are plotted as a function of the percentage of voxels removed. \textbf{(G)}  Mean distance errors are plotted as a function of the percentage of voxels removed. } \label{fig:3}
\end{figure}

We first tested both methods on simulations with voxels with average sides of \SI{5}{\micro\meter} (\autoref{fig:3}A). The SDM method led to a faithful reconstruction, with the exception that the reconstructed voxels tended to be overrepresented around the outside of the cube and underrepresented in the middle. The ULI method also leads to a faithful reconstruction (\autoref{fig:3}B).

We next tested both methods of reconstruction while varying the voxel size (\autoref{fig:3}C,D,E). While voxels could be any shape, for ease of understanding, we report voxel sizes as the edge length of the cube corresponding to the average voxel size. For most voxel sizes, ULI leads to a slightly more accurate reconstruction than SDM. In general, when looking at the error in terms of absolute distance, using smaller voxels increases possible resolution. As an example of the excellent resolution that can be achieved, both methods achieve mean errors below \SI{6}{\micro\meter} using an average voxel size of \SI{3}{\micro\meter} (\autoref{fig:3}E).

Finally, we tested the performance of puzzle imaging when removing a fraction of the voxels. We did this because in real neurons, there are cell bodies that would fill multiple voxels. These voxels would not add any information to puzzle reconstruction, so we exclude them to simulate the existence of cell bodies. Moreover, this generally simulates the robustness of neural voxel puzzling to missing data. In these simulations, we used a voxel size of \SI{5}{\micro\meter} (\autoref{fig:3}F,G). For both methods, when no voxels were removed, the average mean error rate was below 2 voxels. The average mean error rate was still below 2.5 voxels when 60\% of the voxels were removed. This demonstrates that these methods are robust to missing voxels (cell bodies).

\subsection{Neural Connectomics Puzzling} 

\subsubsection{Overview} 
The purpose of neural connectomics puzzling is to estimate the locations of neurons based on their neural connections. There is ongoing work to provide a unique DNA barcode to every neuron and link those barcodes when neurons are synaptically connected~\cite{zador12}. This DNA could later be sequenced to determine all neural connections. For this proposed technology, it would be useful to know the locations of connected neurons, rather than simply knowing that two neurons are connected.

The first steps in connectomics puzzling are to label each neuron in the brain with a unique DNA barcode and label neural connections via the pairing of barcodes (\autoref{fig:4}A). One proposed method would be to have viruses shuttle the barcodes across synapses, where they can be integrated~\cite{zador12}. Other techniques for creating, pairing, and transporting barcodes have been proposed~\cite{marblestone14,marblestone13}. Next, the brain is homogenized, and the DNA barcode pairs are sequenced, yielding a record of which neurons are connected (\autoref{fig:4}B). 

\subsubsection{Dimensionality Reduction} 
\begin{figure}[H]
\centering
\includegraphics[scale=1]{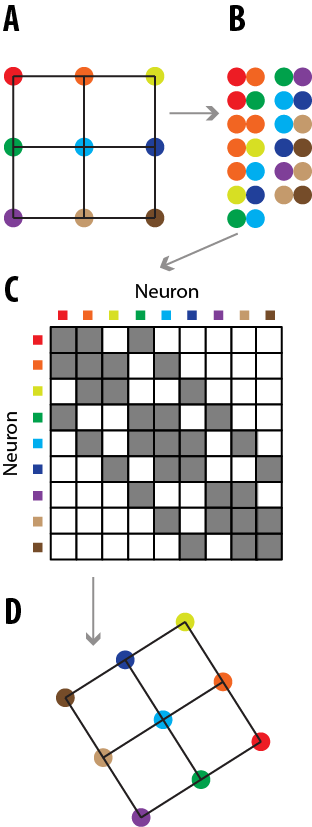}
\caption[Neural Connectomics Puzzling Overview]{\textbf{Neural Connectomics Puzzling Overview.} \textbf{(A)} An example of 9 connected neurons (circles). Lines signify connections. \textbf{(B)} After the brain is homogenized, the only remaining information is a record of the connections. Connections are shown here as adjacent circles. \textbf{(C)} A connectivity matrix is constructed describing the connections between neurons. Gray signifies a connection. Since connections are correlated with how close neurons are to one another, this connectivity matrix can be treated as the similarity matrix. \textbf{(D)} The neurons are puzzled back together. The formation may be rotated or flipped, as shown here.} \label{fig:4}
\end{figure}

Each neuron can be treated as an $N$-dimensional object, where $N$ is the number of neurons. Each dimension corresponds to a connection with a given neuron (\autoref{fig:4}C).  These $N$-dimensional neurons must be mapped to their correct 3-dimensional coordinates (or 2 dimensions in Figure 4's simplified example). This can be done because neurons are more likely to be connected to nearby neurons (e.g.~\cite{hellwig00,chen06}), and thus we can treat the connectivity matrix as a similarity matrix. Using this similarity matrix, we can then puzzle the neurons back into place (\autoref{fig:4}D). As with voxel puzzling, it is impossible to know exact locations without any additional information; only the relative locations of neurons can be known (\autoref{fig:4}D).

In order to convert the knowledge of which neurons are connected (\autoref{fig:4}C) to a reconstructed puzzle (\autoref{fig:4}D), we used SDM, which we previously used for Neural Voxel Puzzling. See \textit{\nameref{subsec:algcomp}} for an explanation on why ULI will not work for Neural Connectomics Puzzling.

\subsubsection{Performance} 
\begin{figure}[b!]
\centering
\includegraphics[scale=.82]{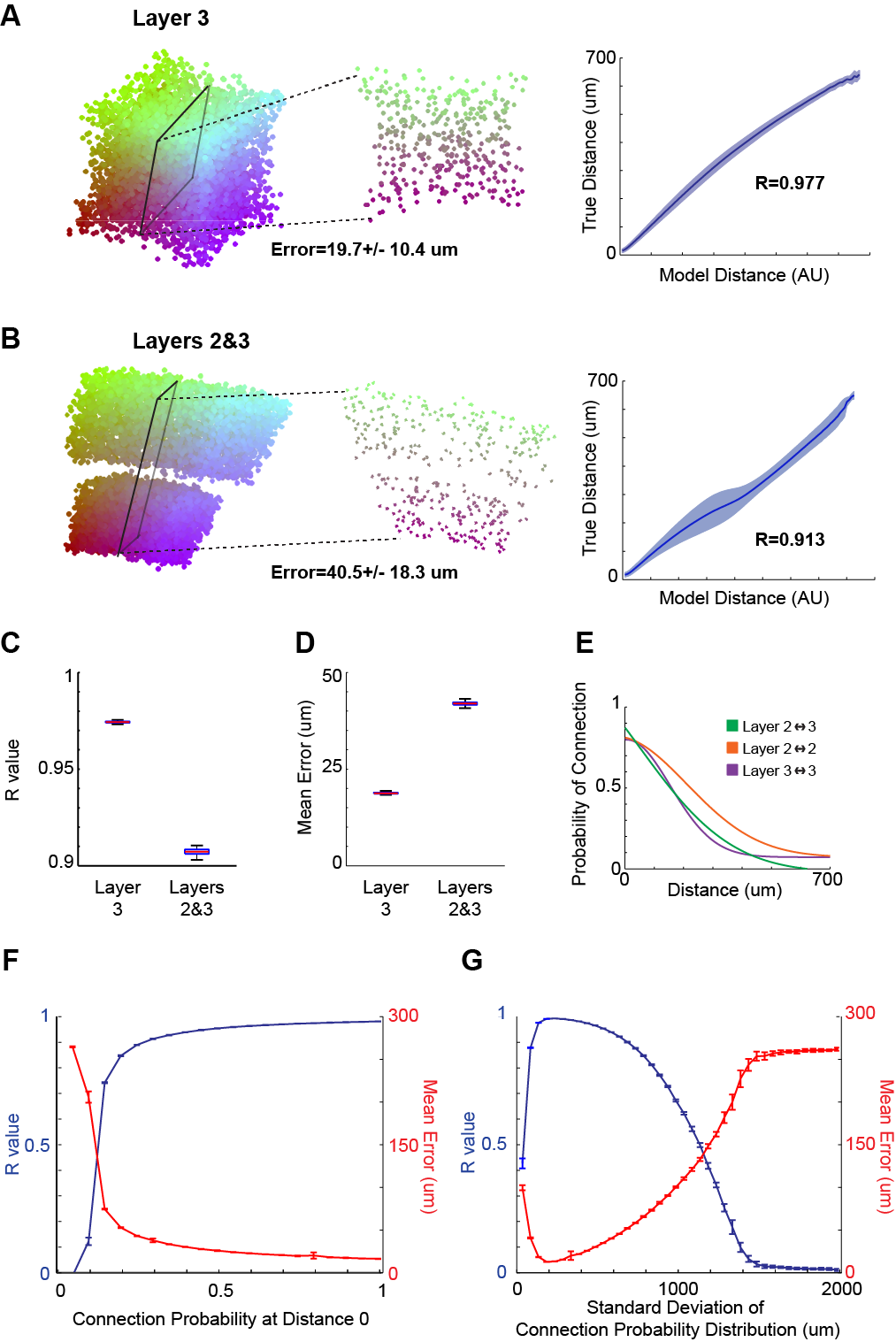}
\caption[Neural Connectomics Puzzling Performance]{\textbf{Neural Connectomics Puzzling Performance.}  } \label{fig:5}
\end{figure}

We tested the ability of connectomics puzzling to determine the locations of a simulated network of neurons. Hellwig~\cite{hellwig00} described the probability of connections as a function of distance between pyramidal cells in layers 2 and 3 of the rat visual cortex. We simulated 8000 neurons in a \SI{400}{\micro\meter} edge-length cube (so they were on average spaced $\sim$\SI{20}{\micro\meter} apart from each other in a given direction). Connections between neurons were randomly determined based on their distance using the previously mentioned probability function (\autoref{fig:5}E). 

We first simulated only connections within layer 3. As our example simulation shows (\autoref{fig:5}A), the locations of neurons were able to be estimated very accurately. As with voxel puzzling, we use mean error and R values to describe the quality of reconstruction, although now the distance is between neurons rather than between voxels. Next we simulated connections between layers 2 and 3; in our simulation, half of the neurons were in layer 2 and half were in layer 3. In an example simulation (\autoref{fig:5}B), it's clear that the locations of neurons could still be estimated accurately. In fact, the reconstruction clearly separated the neurons from the two layers. 

Looking quantitatively at the differences between layer 3 reconstructions and layer 2/3 reconstructions, R values are slightly higher, and mean errors are slightly lower for layer 3 reconstructions. Median R values across simulations are 0.97 vs. 0.91 (layer 3 vs. 2/3), and median mean errors across simulations are \SI{19}{\micro\meter} vs. \SI{42}{\micro\meter} (\autoref{fig:5}C,D). This disparity is largely due to the gap between layers in the reconstructions (\autoref{fig:5}B); reconstructions within each layer are as accurate (in terms of R values) as the layer 3 reconstruction. 

Next, in order to test when connectomics puzzling would be useful, we tested how reconstruction is dependent on the parameters of the connection probability distribution. For these simulations, we assumed the layer 3 connection probability distribution, and then changed either the baseline connection probability (the connection probability at a distance of 0) or the standard deviation of the connection probability distribution. There was a great improvement in reconstruction accuracy when the baseline probability increased from 0.10 to 0.15, and accuracy continued to increase until a baseline probability of 0.35, where it plateaued (\autoref{fig:5}F). In general, there are high accuracy reconstructions for a wide range of baseline probabilities, with the exception of low probabilities. 

When looking at the effect of the standard deviation of the probability distribution, there was a general trend that reconstruction accuracy decreased as the standard deviation increased (\autoref{fig:5}G). This is because connections become less closely related to distances when the standard deviation increases. For example, if the standard deviation was infinite so that the probability of connection was the same for all distances, then knowing the connections would no longer allow us to infer the distances between neurons. 

One notable exception from the above trend is that there is a low reconstruction accuracy for the smallest standard deviations considered (50-\SI{100}{\micro\meter}). We have found that the algorithm does not work well in connectomics puzzling when the connection matrix is far too sparse. When we use a cube with \SI{200}{\micro\meter} edges and 1000 neurons instead of \SI{400}{\micro\meter} edges and 8000 neurons (the same neuronal density), simulations with a \SI{50}{\micro\meter} standard deviation perform very well. Thus, it's important to note that the quality of reconstruction will depend on the size of the cube being used.

\addtocounter{figure}{-1}
\begin{figure} [t!]
  \caption{\textbf{(A)} On the \textbf{left}, an example reconstruction of neural locations based on a simulation of connections of pyramidal cells in layer 3 of cortex. Colors are based on initial locations: those with a larger initial x location are more red, while those with a larger initial y location are more blue. The distance errors are calculated following scaling and rotating the reconstructed volume to match the original volume. In the \textbf{middle}, a 2-dimensional slice through reconstructed volume. On the \textbf{right}, one metric for the accuracy of reconstruction is shown by plotting the reconstructed distances between all points against their true distances for the reconstruction in this panel. The mean plus/minus the standard deviation (shaded) is shown. A perfect reconstruction would be a straight line, corresponding to an R value of 1. \textbf{(B)} Same as panel A, except based on an example simulation of pyramidal cells in both layers 2 and 3 of cortex. \textbf{(C)} Boxplots of R values for layer 3 simulations, and layer 2/3 simulations. The 5 lines (from bottom to top) correspond to 5\%, 25\%, 50\%, 75\%, and 95\% quantiles of R values across simulations. \textbf{(D)} Boxplots (as in panel C) of mean errors across simulations. \textbf{(E)} The probability of connection as a function of distance between pyramidal cells, which is used in the simulations of the other panels~\cite{hellwig00}. \textbf{(F)} Using the parameters of the connectivity probability distribution of layer 3, the baseline connection probability (the probability of connection at a distance of 0) of the connectivity distribution is changed. R values and mean errors are shown as a function of this baseline probability. Error bars represent the standard deviation across simulations in this and the next panel. \textbf{(G)} Using the parameters of the connectivity probability distribution of layer 3, the standard deviation of the connectivity distribution is changed. R values and mean errors are shown as a function of the standard deviation.}
\end{figure}

\subsection{Chemical Puzzling}
\subsubsection{Overview}

Puzzle imaging has a number of potential applications besides neuroscience, including some that are potentially more tractable and therefore may have a higher probability of being implemented in the near future. One example we develop here is that of reconstructing spatial chemical environments from populations of recording cells that have been mixed, a process we call ``chemical puzzling."

One example of a chemical puzzling assay would consist of the initial spreading of many ``pioneer" cells across an environment containing a heterogeneous distribution of a particular chemical (\autoref{fig:6}A,B). These pioneer cells would be endowed with the ability to detect the presence of that chemical and to record its concentration into a nucleotide sequence. This could be done through molecular ticker-tape methods using DNA polymerases~\cite{kording11,zamft12,glaser13}, similar strategies using RNA polymerases, or other mechanisms~\cite{friedland09,church03,farzadfard14,moser13}, for example involving chemically-induced methylation or recombination. 

Some pioneer cells would also be given the ability to share genetic information with other cells --- in the case of prokaryotes, this could be by the introduction of barcoded F-like plasmids encoding the components essential to conjugative transfer of the plasmid~\cite{willetts80}. The pioneer cells would be placed at random places within the chemical environment (\autoref{fig:6}B), and would then begin to grow outwards (\autoref{fig:6}C). When the colonies become large enough to contact neighboring colonies, those that contain the F-plasmid (or equivalent), denoted F\textsuperscript{+}, will copy it and transfer it to those without it (F\textsuperscript{-}) (\autoref{fig:6}D). We can think about the F-plasmid transfer between the colonies descended from two pioneer cells as these two pioneer cells becoming ``connected" (just like a neural connection in the previous section). This sharing of genetic information provides information about which pioneer cells are close to each other, which can then be used to reconstruct where each cell was spatially.

Again, note that a benefit resulting from this strategy is that the spatial information can be destroyed and reconstructed \textit{post hoc}. In terms of this example, this would be equivalent to wiping the surface that the bacteria are growing on and reconstructing its chemical spatial information \textit{after} sequencing, in ``one pot." Preparation of the DNA for sequencing could also be done in one pot, i.e. all of the cells could be lysed and the DNA extracted at one time, if the genetic material carrying the connection information and the chemical information were physically linked (e.g. if a barcode on the F-plasmid were inserted into the chemical record region of the recipient cell's genetic information). Otherwise, amplification of the connection information and the chemical record could be done on each cell using emulsion-based methods~\cite{nakano03}.

\subsubsection{Dimensionality Reduction}

\begin{figure}[H]
\centering
\includegraphics[scale=1]{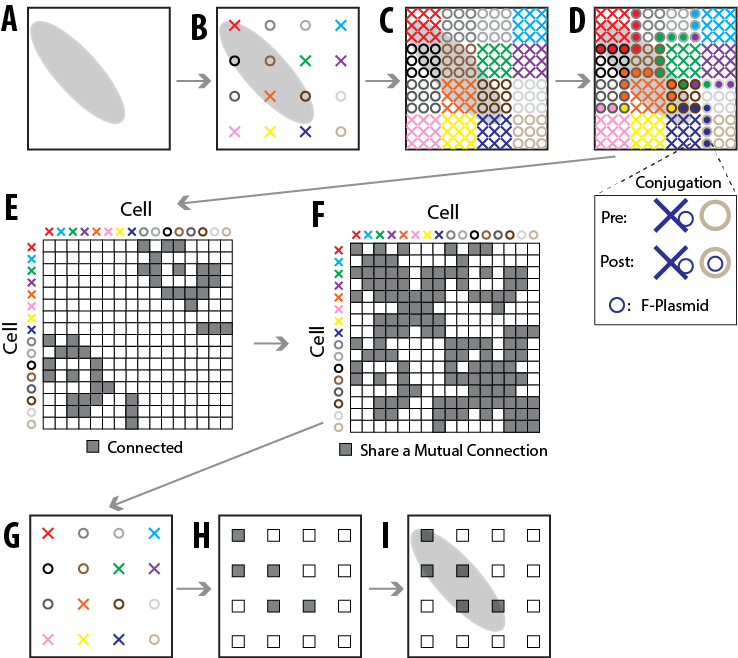}
\caption[Chemical Puzzling Overview]{\textbf{Chemical Puzzling Overview.} \textbf{(A)} An example area contains a chemical whose concentration is represented by a grayscale value. \textbf{(B)} Pioneer cells (X's and O's, each with a different color) are put on the plate. X's represent F\textsuperscript{+} cells that can transfer an F-plasmid into an F\textsuperscript{-} cell (O's). \textbf{(C)} The pioneer cells replicate and spread. Descendants of a pioneer are shown in the same color. \textbf{(D)} When the cell colonies become large enough to contact neighboring colonies, the F\textsuperscript{+} cells (X's) will copy the F-plasmid and transfer it to the F\textsuperscript{-} cells (O's). This is shown as the X's color filling in the center of the O's. In the inset (below), the F-plasmid transfer (conjugation) is shown. \textbf{(E)} The DNA is sequenced to determine which pioneer cells are ``connected" (which had a conjugative transfer occur between their colonies). A connectivity matrix is made from this data. \textbf{(F)} The matrix of connections doesn't directly provide accurate information about how close the original cells are to each other because O's can't be connected to O's (and same for X's).  As our similarity matrix, we thus use the matrix of mutual connections, which allows O's to be connected to O's. \textbf{(G)} The location of the original cells is estimated from the matrix in panel F. \textbf{(H)} The chemical concentrations at each of the original cells locations is known as the cells' DNA acts as a chemical sensor. \textbf{(I)} The chemical concentration everywhere is extrapolated based on the chemical concentrations at the known pioneer cells. } \label{fig:6}
\end{figure}

Each pioneer cell can be thought of as an $N$-dimensional object, where $N$ is the number of pioneer cells. Each dimension corresponds to a connection with another pioneer cell (whether the pioneer cell's descendants are involved in a conjugative transfer with another pioneer cell's descendants) (\autoref{fig:6}E).  These $N$-dimensional cells must be mapped to their correct 2-dimensional coordinates. This can be done because pioneer cells will be more likely to be connected to nearby pioneer cells. We construct a similarity matrix by determining all cells two connections away. This is because the matrix of connections doesn't directly provide accurate information about how close the pioneer cells are to each other, as F\textsuperscript{-}'s can't be connected to F\textsuperscript{-}'s (and same for F\textsuperscript{+}'s).  The matrix of mutual connections allows F\textsuperscript{-}'s to be connected to F\textsuperscript{-}'s. With the knowledge of the similarity between all pairs of pioneer cells, we can puzzle the pioneers back into place (\autoref{fig:6}G). The chemical recording functionality of the cells then allows the chemical environment to be determined at the locations of the pioneer cells (\autoref{fig:6}H), and extrapolated beyond (\autoref{fig:6}I).

In order to convert the knowledge of the similarity between cells (\autoref{fig:6}F) to a reconstructed puzzle (\autoref{fig:6}G), a dimensionality reduction algorithm is needed. Here we use Unweighted Landmark Isomap, as we used in Neural Voxel Puzzling, with only 5 landmark points. See \textit{\nameref{subsec:algcomp}} for an explanation on why SDM will not work well for Chemical Puzzling.

\subsubsection{Performance}

\begin{figure}[H]
\centering
\includegraphics[scale=0.82]{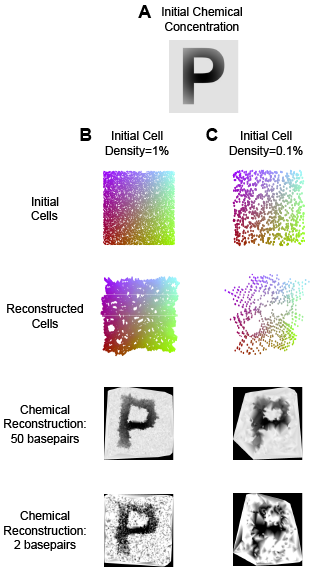}
\caption[Chemical Puzzling Performance]{\textbf{Chemical Puzzling Performance.} \textbf{(A)} The chemical concentration across the plate. It is described by the letter ``P," with the concentration decreasing moving outwards from the center, and a constant background concentration. \textbf{(B)} A simulation is done with an initial cell density of 1\%. \textbf{(C)} A simulation is done with an initial cell density of 0.1\%. For panels B and C, the \textbf{top row} shows the initial locations of the pioneer cells. They are color-coded by location. The \textbf{second row} shows example reconstructed locations of the pioneer cells.  The \textbf{third row} shows the reconstructed chemical concentrations when 50 base pairs are used to detect the concentration. The \textbf{bottom row} shows the reconstructed chemical concentrations when 2 base pairs are used to detect the concentration. Note that the black border represents regions of unknown concentration. } \label{fig:7}
\end{figure}

To further demonstrate the potential for chemical puzzling, we performed a simulation of the chemical puzzling problem. We used a complex chemical concentration described by the letter ``P" (for Puzzle Imaging), with the concentration also decreasing when moving outwards from the center, and a constant background concentration (\autoref{fig:7}A). The image size is 1000 x 1000 pixels, and each pixel is $\SI{1}{\micro\meter}^2$ (about the size of a cell~\cite{grossman82}). The corresponding size of the letter, then, is about \SI{600}{\micro\meter} x \SI{800}{\micro\meter}. Each pioneer cell is randomly placed on a single pixel. 

We simulated the placement and growth of thousands of pioneer cells in that environment (\autoref{fig:7}B,C). Each cell synthesized a DNA element in which the percentage GC composition of the synthesized DNA was proportional to the chemical concentration at that spot. We were able to successfully reconstruct the letter ``P" in the image (\autoref{fig:7}B,C), though this was dependent on the number of pioneer cells used and the length of the incorporated DNA element. Fewer pioneer cells resulted in a decrease in the spatial resolution, whereas the dynamic range greatly increased with longer DNA incorporations (\autoref{fig:7}B,C). When only two base pairs are used, the background concentration and the concentration decrease away from the center were unable to be accurately detected. With 50 base pairs, the chemical concentrations were reconstructed very accurately.

Lastly, the above simulations assumed that when a F\textsuperscript{+} and F\textsuperscript{-} cell are in contact, transfer of genetic information occurs 100\% of the time, which would likely not occur~\cite{ponciano07}. Still, relatively faithful chemical reconstruction was accomplished with conjugation efficiencies as low as 30\% (see \autoref{fig:S1} and \textit{\nameref{sec:sup}}). Overall, this technique holds the potential to determine chemical concentrations at very high resolution.

\section{Discussion} 

Here we proposed the concept of puzzle imaging. We developed two possible large-scale nonlinear dimensionality reduction algorithms for use in puzzle imaging, and demonstrated some of puzzle imaging's abilities and weaknesses in three possible applications. Using simplistic simulations, we showed that voxel puzzling may allow locating neural structures within about \SI{10}{\micro\meter}. In regards to connectomics puzzling, knowing only the connections between neurons within a layer of cortex could be sufficient to localize neurons within about \SI{20}{\micro\meter}. We also showed that chemical puzzling could be used to accurately determine chemical concentrations at a resolution well below 1 mm. Lastly, we describe how Sparse Diffusion Maps is faster than Diffusion Maps, and Unweighted Landmark Isomap is faster than Isomap (see \textit{\nameref{sec:methods}}).

\subsection{Potential Uses of Puzzle Imaging}
For neuroscience, puzzle imaging could be a scalable way to localize important molecular information within the brain. Neuronal RNA/DNA barcodes could be annotated with additional molecular information~\cite{marblestone14,marblestone13}. For example, molecular ticker tapes have been proposed that would record the activity of a neuron into DNA~\cite{kording11,zamft12,glaser13}. It would be very valuable to know the location of the neurons that are being recorded from. Additionally, RNA that is being expressed could be annotated to the barcodes. This could provide information about what genes are being expressed in different locations of the brain, or about the distribution of cell types throughout the brain. If engineered cells capable of recording and/or stimulating adjacent neurons can circulate in and out of the brain, the concepts outlined here might help achieve input/output to many/all neurons without surgery.  

The applications of chemical puzzling go beyond that of determining the chemical composition of bacterial cells growing on a surface. Indeed, biology often has the ability to survive and thrive in the harshest of environments, including spaces too small or desolate for even robotic access. Biological sensing and recording can open these areas to characterization and perhaps utilization. Applications could include those in the fields of geology (in which the composition of fractured geologic strata, which contain millions of microscopic cracks and pores, can be assayed) and medicine (in which the composition of the complex biological environments of, for example, the gastrointestinal tract, can be assayed).

\subsection{Simulation Limitations}
In all our simulations, we made simplifications in order to provide a preliminary demonstration of the feasibility of puzzle imaging. In neural voxel puzzling, our simulations assumed that there were equal neuronal densities across the volume and that neurons were oriented at random angles. In neural connectomics puzzling, our simulations used a maximum of two neuronal layers, and assumed that connection probability distributions did not differ within a layer (although in reality there are different cell types~\cite{liley94}). For both of these neuroscience examples, reconstruction errors due to these simplifying assumptions could likely be remedied by using previously known neuroanatomical information (e.g.  the orientation of axons in a brain region) or molecular annotations (e.g. about whether a barcode is in an axon or dendrite, or about cell type~\cite{usoskin14}). In chemical puzzling, our simulations assumed that cells were stationary, and we used a simple model of outward cellular growth.  For locations with viscous flow or rapidly moving cells, complex cell growth and movement models would be necessary to achieve accurate reconstructions. 

For a more in-depth discussion of limitations for all simulations, see \textit{\nameref{sec:sup}}.

\subsection{Algorithms}
In this paper, we demonstrated two algorithms that could be used for puzzle imaging. Refinements of the presented algorithms would be beneficial for puzzle imaging (to overcome the limitations in \textit{\nameref{sec:results}}). For instance, a different metric could be used to create the similarity matrix in both methods. In addition, the number of landmark points in Unweighted Landmark Isomap can be optimally tuned for the specific use. Moreover, novel algorithms for large-scale nonlinear dimensionality reduction would be beneficial for more accurate puzzle imaging reconstruction.  

While the algorithms presented here were designed for puzzle imaging, they could be generally used as faster versions of Diffusion Maps and Landmark Isomap for large problems. Both methods can preserve relative locations when reconstructing a swiss roll or helix, classical tests of nonlinear dimensionality reduction techniques. Further research needs to be done to see how these methods compare to traditional methods for other applications.

\subsection{Conclusion}
Here we have provided a preliminary demonstration that puzzle imaging may be possible. In order to make puzzle imaging a reality, significant biological work still needs to be done. For example, having location or neuron specific barcodes would be needed to make the neuroscience puzzle imaging approaches possible. We hope that this paper will inspire experimentalists and theoreticians to collaborate to help make puzzle imaging a reality.

\section{Methods} 
\label{sec:methods}

\subsection{Sparse Diffusion Maps Algorithm}

Sparse Diffusion Maps is the same as the standard Diffusion maps algorithm~\cite{coifman06} when using 1 timestep, except we here use a sparse similarity matrix. Thus, in the below algorithm, only step 1 differs from standard Diffusion maps. The algorithms take $N$ high-dimensional points, and reduce each point to $k$ dimensions (where $k$ is 2 or 3 in our applications). The algorithm follows in \autoref{tab:1}.

\begin{table} [h]
\centering
\begin{tabular}{| l | p{15 cm} |}
\hline

\multicolumn{2}{ |c| }{\textbf{Sparse Diffusion Maps Algorithm}} \\
\hline
Step 1 & Efficiently create a sparse, symmetric, nonnegative, similarity matrix, $\mathbf{S}$ that is a connected graph. 
\newline Methods used to create $\mathbf{S}$ in our applications follow. \\ \hline
Step 2 & Create matrix $\mathbf{M}$ by normalizing $\mathbf{S}$ so that each row sums to 1. That is, $\mathbf{M}={{\mathbf{D}}^{-1}}\mathbf{S}$, where $\mathbf{D}$ is a diagonal matrix with ${{\mathbf{D}}_{i,i}}=\sum\limits_{j}{{{\mathbf{S}}_{i,j}}}$ \\ \hline
Step 3 & Find the $k+1$ largest eigenvalues ${{e}_{1}},...,{{e}_{k+1}}$ of $\mathbf{M}$ and their corresponding eigenvectors ${{\mathbf{v}}_{1}},...,{{\mathbf{v}}_{k+1}}$. Each eigenvector is $N$-dimensional.  \\ \hline
Step 4 & The final $k$-dimensional positions of the $N$ points are the $N$ rows in $\left[ {{e}_{2}}{{\mathbf{v}}_{2}},...,{{e}_{k+1}}{{\mathbf{v}}_{k+1}} \right]$. \\ \hline
\end{tabular}
\caption[Sparse Diffusion Maps Algorithm]{\textbf{Sparse Diffusion Maps Algorithm}}
\label{tab:1}
\end{table}

\subsection{Unweighted Landmark Isomap Algorithm}
\label{subsec:uli}

Unweighted Landmark Isomap is based on Landmark Isomap~\cite{desilva04,silva02}. The most important change is that we compute geodesic distances more efficiently due to our graph being unweighted (Step 3). Additionally, we create the similarity matrices uniquely for each application (Step 1). All other steps are identical to Landmark Isomap. The algorithm follows in \autoref{tab:2}.

\begin{table} [h]
\centering
\begin{tabular}{| l | p{15 cm} |}
\hline

\multicolumn{2}{ |c| }{\textbf{Unweighted Landmark Isomap Algorithm}} \\
\hline
Step 1 & Efficiently create a sparse, binary, symmetric, similarity matrix, $\mathbf{S}$ that is a connected graph.
\newline Methods used to create $\mathbf{S}$ in our applications follow. \\ \hline
Steps 2\&3 & These steps select a landmark point (Step 2) and then calculate the distance from all points to that landmark point (Step 3). In total, we will select $\ell $ landmark points, so these steps will be repeated in alternation $\ell$ times. \\ \hline
Step 2 & Select a landmark point.  We do this in the following way (MaxMin algorithm  from~\cite{desilva04}): 
\newline The first landmark point is selected randomly.
Each additional landmark point is chosen in succession in order to maximize the minimum distance to the previously selected landmark points. In other words, for every point $i$, we calculate the minimum distance ${{m}_{i}}$ to all the previously selected landmark points (indexed by $j$): ${{m}_{i}}=\underset{j}{\mathop{\min }}\,{{\gamma }_{i,j}}$, where ${{\gamma }_{i,j}}$ is the calculated geodesic distance between the points in Step 3. We then choose the point that has the largest minimum distance: $landmark=\underset{i}{\mathop{\max }}\,{{m}_{i}}$.
 \\ \hline
Step 3 & Calculate the geodesic distance from all $N$ points to the previously selected landmark point. The geodesic distance ${{\gamma }_{i,j}}$ from point $i$ to landmark point $j$ is the number of steps it would to get from point $i$ to landmark point $j$ in the graph of $\mathbf{S}$. 	This can be calculated efficiently using a breadth first search (as opposed to using Dijkstra's algorithm) because $\mathbf{S}$ is an unweighted graph.   \\ \hline
Step 4 & Use classical multidimensional scaling (MDS)~\cite{kruskal64} to place the landmark points. This means that the points are placed in order to minimize the sum squared error between the Euclidean distances of the placed landmark points and their known geodesic distances. 
This is done in the following way~\cite{desilva04,silva02}:

Let $\mathbf{\Delta^\ell} $ be the matrix of squared geodesic distances between all the landmark points. 
Let ${{\mu }_{i}}$ be the mean of row $i$ of $\mathbf{\Delta^\ell} $ and $\mu $ be the mean of all entries of $\mathbf{\Delta^\ell} $.
Create the matrix $\mathbf{B}$ with entries ${{\mathbf{B}}_{j,k}}=[\Delta^\ell]_{j,k} -{{\mu }_{j}}-{{\mu }_{k}}+\mu $.
Find the first (largest) $k$ eigenvalues and eigenvectors of $\mathbf{B}$. Let ${{\mathbf{V}}_{i,j}}$ be the $i^{th}$ component of the $j^{th}$ eigenvector. Let ${{\lambda }_{j}}$ be the $j^{th}$ eigenvalue.
The matrix $\mathbf{L}$ has the coordinates of the landmark points in its columns. $\mathbf{L}$ has the entries ${{\mathbf{L}}_{i,j}}={{\mathbf{V}}_{j,i}}\cdot \sqrt{{{\lambda }_{i}}}$.
 \\ \hline
Step 5 & For each point, triangulate its location using the known geodesic distances to each of the landmark points. This is done in the following way~\cite{desilva04,silva02}:

Let ${{\mathbf{L}}^{\#}}_{i,j}=\frac{{{\mathbf{V}}_{j,i}}}{\sqrt{{{\lambda }_{i}}}}$, and let $\mathbf{\Delta} $ be the matrix of squared geodesic distances between the landmark points and all points.
The position of point $a$, ${{\mathbf{x}}_{a}}$, is calculated by ${{\mathbf{x}}_{a}}=-\frac{1}{2}\mathbf{L^{\#}}\left( {{\mathbf{\Delta }_{a}}}-\mathbf{\bar\Delta^\ell}  \right)$, where ${\mathbf{{\Delta }}_{a}}$ is row $a$ of the matrix $\mathbf{\Delta} $ (the squared geodesic distances from point $a$ to all the landmark points), and $ \mathbf{\bar\Delta^\ell} $ is the column mean of the matrix $ \mathbf{\Delta^\ell} $. \\ \hline
\end{tabular}
\caption[Unweighted Landmark Isomap Algorithm]{\textbf{Unweighted Landmark Isomap Algorithm}}
\label{tab:2}
\end{table}

Note that the method of constructing the similarity matrix (Step 1) and the number of landmark points (Step 2) are user options within this algorithm. We discuss our choices for these steps for our applications below.

\subsection{Neural Voxel Puzzling}
\subsubsection{Similarity Matrix}

Let $\mathbf{X}$ be an $n\times p$ (voxels $\times $ neurons) coincidence matrix that states which neurons are in which voxel. To construct the similarity matrix, we first compute $\mathbf{A=X}{{\mathbf{X}}^{T}}$, an $n\times n$ matrix that gives the similarity between voxels. We then threshold this matrix (an approximate nearest neighbors calculation). Thresholding makes the matrix more sparse (beneficial for SDM), and makes points not connected to far-away points (important for increasing resolution in ULI). We use different thresholding methods prior to ULI and SDM. The specific thresholding methods below are what we used for our simulations; the method of thresholding is a tunable parameter.

For ULI, we threshold each row independently. The threshold for a row is the maximum of that row divided by 2. That is, for row $i$, ${{T}_{i}}=\frac{1}{2}\underset{j}{\mathop{\max }}\,{{\mathbf{A}}_{ij}}$, where ${{\mathbf{A}}_{ij}}$ is the element of $\mathbf{A}$ in row $i$ and column $j$. After thresholding, the entries of the non-symmetric similarity matrix are ${{\mathbf{N}}_{ij}}={{\mathbf{A}}_{ij}}\ge {{T}_{i}}$. We create a symmetric unweighted similarity matrix: $\mathbf{S}=\left( \mathbf{N}+{{\mathbf{N}}^{T}} \right)>0$.

For SDM, the entire matrix is thresholded by the same value. We first create a vector $\mathbf{w}$ that contains the maximum of each row of $\mathbf{A}$. Its entries are ${{w}_{i}}=\underset{j}{\mathop{\max }}\,{{\mathbf{A}}_{ij}}$. The threshold is the minimum of $\mathbf{w}$. That is, $T=\min \mathbf{w}$. Thresholding all entries, we get $\mathbf{S}=\mathbf{A}\ge T$.

\subsubsection{Simulations}

For our simulations, we initially chose the average voxel size. If the average voxel had a volume of $x^3$ $\SI{}{\micro\meter}^3$, we reported the voxel size as $x$ \SI{}{\micro\meter} (the length of an edge of a cube of comparable size). We next created a cube with edges of length $20x$ \SI{}{\micro\meter} (so it had a volume of $8000x^3$ $\SI{}{\micro\meter}^3$). We then randomly placed 8000 voxel centers in the cube. We next added in long rectangular prisms (neurons) so that the entire volume would be filled. We assumed the neurons fully went through each voxel they entered. As the cross-sectional area of a voxel was on average $x^2$ $\SI{}{\micro\meter}^2$, and we assumed neurons with cross-sectional areas of $\SI{1}{\micro\meter}^2$ (about the size of an axon), we added enough neurons so that each voxel would contain $x^2$ neurons. Neurons were oriented within the volume at randomly determined angles.

When doing simulations with removed voxels, we placed voxels and neurons in the regular way. Then, we discarded voxels and aimed to reconstruct the locations of the other voxels. The R values and mean errors listed are for the voxels that were reconstructed (removed voxels were not included in the analysis).

Additionally, in our simulations, voxels that did not contain any neurons, or that did not share any neurons with another voxel, were excluded from analysis. When voxels were \SI{2}{\micro\meter}, \SI{3}{\micro\meter}, \SI{4}{\micro\meter}, and \SI{5}{\micro\meter}, 2.8\%, 0.14\%, 0.0046\%, and 0.0003\% of voxels were respectively excluded. Also, when removing voxels (\autoref{fig:2}F,G), we excluded the remaining voxels that did not make up the main connected component of the similarity graph. When using ULI, when 80\%, 85\%, and 90\% of voxels were removed, 0.36\%, 1.7\%, and 34.7\% of remaining voxels were excluded from analysis. When using SDM, when 80\%, 85\%, and 90\% of voxels were removed, 0.05\%, 0.09\%, and 0.37\% of remaining voxels were excluded from analysis.

In our simulations, when using ULI, we used 10 landmark points (see \textit{\nameref{subsec:uli}} above).

\subsection{Neural Connectomics Puzzling}
\subsubsection{Similarity Matrix}
Let $\mathbf{C}$ be an $n\times n$ (neurons $\times $ neurons) connectivity matrix that states which neurons are connected to each other. As mentioned in \textit{\nameref{sec:results}}, this connectivity matrix can be directly used as the similarity matrix.

\subsubsection{Simulations}

We randomly placed 8000 points (neurons) in a \SI{400}{\micro\meter} edge-length cube. Thus, the neurons were on average $\sim$\SI{20}{\micro\meter} apart from each other in a single direction. We initially assumed all neurons were pyramidal cells in layer 3 of rat visual cortex and simulated connections according to the known connection probability distribution as a function of distance (\autoref{fig:5}E)~\cite{hellwig00}. Every pair of neurons was randomly assigned a connection (or not) based on this probability distribution. Next, we simulated neurons in layers 2 and 3. The top half of the cube was assumed to be layer 2, and the bottom half was assumed to be layer 3. Again, every pair of neurons was randomly assigned a connection (or not) based on the relevant probability distribution (either between layer 2 and 2, layer 2 and 3, or layer 3 and 3; \autoref{fig:5}E)~\cite{hellwig00}.

\subsection{Chemical Puzzling}
\subsubsection{Similarity Matrix}

Let $\mathbf{C}$ be an $n\times n$ (pioneer cells $\times $ pioneer cells) connectivity matrix that states which pioneer cells are ``connected" to each other. The similarity matrix is calculated as ${{\mathbf{C}}^{T}}\mathbf{C}$.

\subsubsection{Simulations}

We randomly placed pioneer cells on a pixel in the $1000 \times 1000$ pixel image. The number of pioneer cells was 1 million (the number of pixels) times the initial density, so 10000 in \autoref{fig:3}B, and 1000 in \autoref{fig:3}C. Each cell was randomly assigned to be F\textsuperscript{+} or F\textsuperscript{-}. We assumed the cells grew out circularly over time with approximately the same rate of growth. For every pixel, we determined which colony (progeny from which pioneer cell) would reach the pixel first. This produced a final map of all pixels assigned to different pioneer cells. On the borders of colonies, cells could conjugate with one another if one colony was F\textsuperscript{+} and the other was F\textsuperscript{-}. For each pixel on the border, a conjugation occurred according to the probability of conjugation (100\% probability for \autoref{fig:7}; varying for \autoref{fig:S1}). More specifically, any time different colonies occupied pixels horizontal or vertical of each other, a conjugation could occur for the cells in the bordering pixels. 

We reconstructed the pioneer cells' locations using ULI with 5 landmark points (see \textit{\nameref{subsec:uli}} above). We assumed that 3 cells were placed on the plate so that their locations were known. We scaled, rotated, and reflected the reconstructed locations of the pioneer cells so that the locations of the 3 known cells were matched (the average distance error was minimized). 

The chemical concentration on the plate was between 0 and 1. If a pioneer cell was at a location with a chemical concentration of  $\rho$, each base pair has a probability of being a G or C with probability $\rho $. The reconstructed chemical concentration was the total number of G's and C's divided by the number of base pairs. Thus, if there were 2 base pairs, the reconstructed chemical concentration could be 0, 0.5, or 1. We used linear interpolation to determine the chemical concentration of areas on the plate that did not have a reconstructed concentration.

We also note that along with our approximate simulation method of outward circular growth, we also ran a smaller, more realistic, simulation. In this stochastic growth simulation, during each round, cells randomly divided into any single adjacent unoccupied pixel. When an F\textsuperscript{+} colony grew next to an F\textsuperscript{-} colony, or vice versa, cells could conjugate (according to the conjugation probability). F\textsuperscript{-} cells turned F\textsuperscript{+} upon receipt of the F-plasmid from an F\textsuperscript{+} cell. This growth simulation continued until the plate was full. On smaller plates (when it was less time consuming to run the realistic simulation), both simulation types produced nearly equivalent results.

\subsection{Computational Complexity}
\label{subsec:compcomp}

\subsubsection{Sparse Diffusion Maps}
We list the complexity of Step 1 for the individual methods below. The run-time of this algorithm is dominated by Step 3. For the below complexity explanation, $\mathbf{S}$ is an $n\times n$ sparse matrix with $m$ non-zero elements. The complexity of this algorithm is listed in \autoref{tab:3}.

\begin{table} [h]
\centering
\begin{tabular}{| p{1.75 cm} | p{3.5 cm} | p{9.5 cm} |}
\hline

\multicolumn{3}{ |c| }{\textbf{Sparse Diffusion Maps Computational Complexity}} \\
\hline
\textbf{Step} & \textbf{Worst-case complexity} & \textbf{Further Explanation}  \\ \hline
Step 3 \newline(and Total) & $\mathcal{O}\left( m \right)$ & This is the complexity for the power iteration algorithm~\cite{golub12}, which is very efficient for solving for a small number of eigenvalues/eigenvectors. In the power iteration method, the matrix is continually multiplied by a vector, which has complexity $\mathcal{O}\left( m \right)$, until convergence. The number of steps to calculate eigenvalue ${{e}_{i}}$ depends on $\left| {{e}_{i}}/{{e}_{i+1}} \right|$: a larger ratio means quicker convergence.
\newline
\newline
If this was not a sparse matrix, as is the case for standard Diffusion Maps, then this step would be $\mathcal{O}\left( {{n}^{2}} \right)$.
  \\ \hline

\end{tabular}
\caption[Sparse Diffusion Maps Computational Complexity]{\textbf{Sparse Diffusion Maps Computational Complexity}}
\label{tab:3}
\end{table}

\subsubsection{Unweighted Landmark Isomap}
We list the complexity of Step 1 for the individual methods below. The complexity of this algorithm is listed in \autoref{tab:4}.

\begin{table} [h]
\centering
\begin{tabular}{| p{1.75 cm} | p{3.5 cm} | p{9.5 cm} |}
\hline

\multicolumn{3}{ |c| }{\textbf{Unweighted Landmark Isomap Computational Complexity}} \\
\hline
\textbf{Step} & \textbf{Worst-case} \newline \textbf{complexity} & \textbf{Further Explanation}  \\ \hline
Step 2 & $\mathcal{O}\left( \ell \cdot n \right)$ & ~\cite{desilva04}
  \\ \hline
Step 3 & $\mathcal{O}\left( \ell \cdot m \right)$ & A breadth first search has a complexity of $\mathcal{O}\left( m \right)$. This is done for each of $\ell $ landmark points.
\newline
\newline
If the graph had been weighted (as assumed in standard Landmark Isomap), then we would need to use Dijkstra's algorithm. The fastest general implementation is $\mathcal{O}\left( m+n\cdot \log n \right)$~\cite{fredman87} for each landmark point, yielding $\mathcal{O}\left( \ell \cdot m+\ell \cdot n\cdot \log n \right)$.
  \\ \hline
Step 4 & $\mathcal{O}\left( {{\ell }^{3}} \right)$ & ~\cite{desilva04}
  \\ \hline
Step 5 & $\mathcal{O}\left( k\cdot \ell \cdot n \right)$ & ~\cite{desilva04}
  \\ \hline
Total & $\mathcal{O}\left( \ell \cdot m+k\cdot \ell \cdot n+{{\ell }^{3}} \right)$ &
  \\ \hline

\end{tabular}
\caption[Unweighted Landmark Isomap Computational Complexity]{\textbf{Unweighted Landmark Isomap Computational Complexity}}
\label{tab:4}
\end{table}

\subsubsection{Neural Voxel Puzzling}
To compute the similarity matrix, we can take a shortcut instead of performing the sparse matrix multiplication $\mathbf{X}{{\mathbf{X}}^{T}}$ (which will be $\mathcal{O}\left( {{n}^{2}} \right)$ in the best case scenario~\cite{kaplan06}). The entries of $\mathbf{X}{{\mathbf{X}}^{T}}$ tell us how many neurons are shared between particular voxels. This can also be calculated by drawing the bipartite graph that $\mathbf{X}$ describes, with voxels on one side and neurons on the other (and edges when a neuron goes through a voxel). For a given voxel, to determine which voxels it shares a neuron with, we only need to trace all the paths to its connected neurons, and then the paths from those neurons back to voxels. This has a complexity of $\mathcal{O}\left( {{\alpha }_{in}}\cdot {{\alpha }_{out}} \right)$, where ${{\alpha }_{in}}$ is the largest number of neurons in a voxel, and ${{\alpha }_{out}}$ is the largest number of voxels a neuron goes through. Thus, for all voxels, the complexity is $\mathcal{O}\left( n\cdot {{\alpha }_{in}}\cdot {{\alpha }_{out}} \right)$. Determining the similarity matrix can take a comparable amount of time to the steps in ULI or SDM.

Note that this step is faster than the standard method of computing a similarity matrix in Diffusion Maps: ${{\mathbf{S}}_{i,j}}=\exp \left( \frac{-{{\left\| {{x}_{i}}-{{x}_{j}} \right\|}_{2}}^{2}}{2{{\sigma }^{2}}} \right)$, where ${{x}_{i}}$ and ${{x}_{j}}$ are columns within $\mathbf{X}$.  For our sparse matrices, this would take $\mathcal{O}\left( q\cdot n \right)$, where $q$ is the number of nonzero entries in $\mathbf{X}$, due to the pairwise vector subtraction. The Landmark Isomap algorithm does not give a method for computing the similarity matrix.

\subsubsection{Neural Connectomics Puzzling}
As no computation is required to construct the similarity matrix, the overall complexity of Neural Connectomics Puzzling is the complexity of the SDM algorithm.

\subsubsection{Chemical Puzzling}
As in neural voxel puzzling, we can take a shortcut to calculate the similarity matrix instead of performing the sparse matrix multiplication ${{\mathbf{C}}^{T}}\mathbf{C}$ (which would take $\mathcal{O}\left( {{n}^{2}} \right)$ at best for very sparse matrices~\cite{yuster05}). Again, we can construct a bipartite graph representing $\mathbf{C}$, where the pioneer cells are now on both sides of the graph (and there's an edge between connected cells). We can determine if a pioneer cell is connected within 2 steps of another pioneer cell by counting all the ways to get to the other side of the graph (via connections) and back to cells on the same side. For a given cell, this has the complexity $\mathcal{O}\left( {{\beta }^{2}} \right)$, where $\beta $ is the largest number of connections a pioneer cell has (how many different cell colonies that pioneer cell's colony has conjugated with). Thus, the total complexity of determining the similarity matrix is $\mathcal{O}\left( n\cdot {{\beta }^{2}} \right)$.  Determining the similarity matrix can take a comparable amount of time to the steps in ULI.

Additionally, chemical puzzling has the extra step of reconstructing the image of chemical concentrations. The chemical concentrations are known at the reconstructed locations of the pioneer cells, but interpolation needs to be used to estimate the concentrations at other locations. This interpolation step can dominate the total time, depending on the desired resolution and interpolation method.

\subsection{Algorithm Comparison: ULI vs SDM}
\label{subsec:algcomp}
There are pros and cons of both Sparse Diffusion Maps (SDM) and Unweighted Landmark Isomap (ULI), which make them suitable for different applications.

\subsubsection{Domain}
To run ULI, the graph entered into the algorithm must contain only short-range connections. Practically, ULI will not work when there is high similarity (a large value in the similarity matrix) between points that are far away from one another. This is because having high similarity between far away points would make those far away points near each other in the reconstruction, a problem known as ``short circuiting"~\cite{vandermaaten09}. This limitation means that neural connectomics puzzling, which contains long-range connections that are indistinguishable for short-range connections, is not compatible with ULI. SDM, on the other hand, is robust to high similarity values between far-away points, as long as similarity values are generally higher between nearby points. Thus, SDM does work for neural connectomics puzzling.

\subsubsection{Reconstruction Accuracy}
Reconstructions using SDM are generally biased towards having points around the perimeter, and therefore don't faithfully reconstruct the center of the volume. This perimeter bias makes the SDM method unsuitable for use in Chemical puzzling. When used with Chemical puzzling, SDM leads to faulty chemical reconstructions near the center of the image. For neural voxel puzzling, as seen in our simulations (\autoref{fig:3}), ULI was slightly more accurate than SDM, except for when at least 85\% of the voxels had been removed.

Another important note is that the SDM method will only accurately reconstruct the volume when the volume is a cube (\autoref{fig:S2}). Thus, for the SDM method to be used in practice with neural voxel or connectomics puzzling, the brain would need to first be sectioned into cubes, and then the cubes would need to be pieced together. The ULI method, on the other hand has the benefit of accurately reconstructing volumes that are not cubes (\autoref{fig:S2}).

\subsubsection{Speed}
Both SDM and ULI are designed to be fast dimensionality reduction methods for use with large datasets. In practice, when directly comparing their speed in the neural voxel puzzling simulations using 8000 \SI{5}{\micro\meter} voxels, SDM took about 1.5 seconds, while ULI (with 10 landmark points) took about 0.9 seconds on a 2.3 GHz processor running Matlab. The previous times did not include constructing the similarity matrix. See \textit{\nameref{subsec:compcomp}} above, and \textit{\nameref{subsec:algtime}} below, for the computational complexity of larger problems.

\begin{table} [h]
\centering
\begin{tabular}{| p{7.5 cm} | p{7.5 cm} |}
\hline
\multicolumn{2}{ |c| }{\textbf{\textcolor{green}{Pros}/\textcolor{red}{Cons} Summary}} \\
\hline
\textbf{Sparse Diffusion Maps} & \textbf{Unweighted Landmark Isomap}  \\ \hline
\cellcolor{red}Only accurately reconstructs cubes & \cellcolor{green}Can accurately reconstruct non-cubes
  \\ \hline
\cellcolor{red}Biased towards exterior points & \cellcolor{green}Not biased towards exterior points
  \\ \hline
\cellcolor{green}Is robust to problems that have high similarity between some far-away points. & \cellcolor{red}Is generally not robust to problems that have high similarity between some far-away points.
  \\ \hline

\end{tabular}
\caption[Algorithm Comparison Summary: ULI vs. SDM]{\textbf{Algorithm Comparison Summary: ULI vs. SDM}}
\label{tab:5}
\end{table}

\subsection{Algorithm Time Improvements}
\label{subsec:algtime}

Our algorithms took previous algorithms and adapted them to be faster for large-scale puzzle imaging.

\subsubsection{Sparse Diffusion Maps vs. Diffusion Maps}
The difference between Sparse Diffusion Maps and Diffusion Maps is that we construct a sparse similarity matrix. The most time consuming step in these algorithms (besides constructing the similarity matrix) is finding the largest $k$ eigenvalues and eigenvectors, where $k$ is the dimension size you're projecting into (2 or 3 for puzzle imaging). This computation will be significantly faster when the similarity matrix is sparse. For instance, one fast algorithm, ``power iteration,"~\cite{golub12} has a complexity of $\mathcal{O}\left( m \right)$, where $m$ is the number of non-zero elements in the similarity matrix $\mathbf{S}$ (see \textit{\nameref{subsec:compcomp}}). 

Computing the similarity matrix is the other time consuming step in Diffusion Maps. 
Let's say $\mathbf{X}$ is an $n\times p$ matrix, and we want to compute the $n\times n$ similarity matrix $\mathbf{S}$ that gives the similarity between the columns of $\mathbf{X}$. This is generally calculated as ${{\mathbf{S}}_{i,j}}=\exp \left( \frac{-{{\left\| {{x}_{i}}-{{x}_{j}} \right\|}_{2}}^{2}}{2{{\sigma }^{2}}} \right)$, where ${{x}_{i}}$ and ${{x}_{j}}$ are columns within $\mathbf{X}$. Computing $\mathbf{S}$ would have a complexity of $\mathcal{O}\left( q\cdot n \right)$, where $q$ is the number of nonzero entries in $\mathbf{X}$ (see \textit{\nameref{subsec:compcomp}}).  

In our scenarios where we use SDM, we compute a sparse similarity matrix more efficiently. For neural connectomics puzzling, computing the similarity matrix takes no time, as it simply describes the connections. For neural voxel puzzling, $\mathbf{X}{{\mathbf{X}}^{T}}$ (the main step of computing the similarity matrix) can be calculated with a complexity of $\mathcal{O}\left( n\cdot {{\alpha }_{in}}\cdot {{\alpha }_{out}} \right)$, where ${{\alpha }_{in}}$ is the largest number of neurons in a voxel, and ${{\alpha }_{out}}$ is the largest number of voxels a neuron goes through (see \textit{\nameref{subsec:compcomp}}). This approach is thus significantly faster than Diffusion Maps for our problem.

\subsubsection{Unweighted Landmark Isomap vs. Landmark Isomap}

The main difference between ULI and Landmark Isomap is that ULI uses an unweighted similarity matrix. One of the time-consuming steps in Landmark Isomap is computing the geodesic distance between all points and each of the $\ell $ landmark points. For weighted graphs, this can be solved fastest with Dijkstra's algorithm in $\mathcal{O}\left( \ell \cdot m+\ell \cdot n\cdot \log n \right)$~\cite{fredman87}. For unweighted graphs, we can simply use a breadth first search for each landmark point, which has a complexity of $\mathcal{O}\left( \ell \cdot m \right)$. When we include the other steps of the algorithm (other than constructing the similarity matrix), Landmark Isomap has a complexity of $\mathcal{O}\left( \ell \cdot m+\ell \cdot n\cdot \log n+k\cdot \ell \cdot n+{{\ell }^{3}} \right)$, while ULI is faster, with a complexity of $\mathcal{O}\left( \ell \cdot m+k\cdot \ell \cdot n+{{\ell }^{3}} \right)$ (see \textit{\nameref{subsec:compcomp}}).

It is important to note that computing the similarity matrices in neural voxel puzzling and chemical puzzling may take a comparable amount of time as ULI or Landmark Isomap. As mentioned above, computing the similarity matrix for neural voxel puzzling has a complexity of $\mathcal{O}\left( n\cdot {{\alpha }_{in}}\cdot {{\alpha }_{out}} \right)$. For chemical puzzling, computing the similarity matrix has a complexity of $\mathcal{O}\left( n\cdot {{\beta }^{2}} \right)$, where $\beta $ is the largest number of connections a pioneer cell has (how many different cell colonies that pioneer cell's colony has conjugated with). Additionally, for chemical puzzling, using interpolation to estimate the chemical concentration can be a very time-intensive step depending on the resolution desired.

\section{Acknowledgements}
We would like to thank Ted Cybulski and Mohammad Gheshlaghi Azar for very helpful comments on previous versions of the manuscript. We would like to thank Josh Vogelstein for helpful initial discussions. 

Joshua Glaser was supported by NIH grant 5R01MH103910 and NIH grant T32 HD057845. Bradley Zamft was supported by NIH grant 5R01MH103910. George Church acknowledges support from the Office of Naval Research and the NIH Centers of Excellence in Genomic Science. Konrad Kording is funded in part by the Chicago Biomedical Consortium with support from the Searle Funds at The Chicago Community Trust, and is also supported by NIH grants 5R01NS063399, P01NS044393, and 1R01NS074044.

\beginsupplement
\section{Supplementary Information}
\label{sec:sup}

\subsection{Chemical Puzzling Conjugation Efficiency}

\begin{figure}[H]
\centering
\includegraphics[scale=.82]{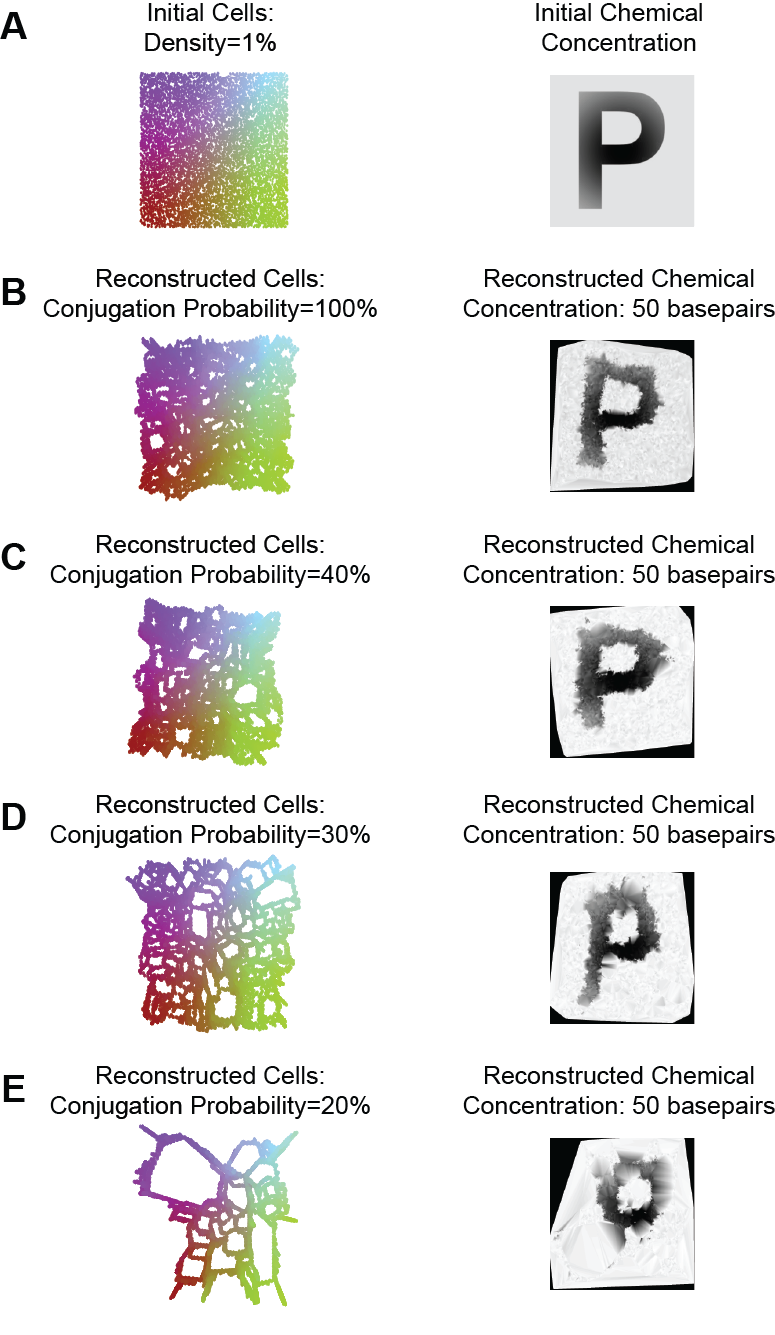}
\caption[Chemical Puzzling with Low Conjugation Efficiency]{\textbf{Chemical Puzzling with Low Conjugation Efficiency.} We do a simulation of chemical puzzling, as in \autoref{fig:7}B, except now with varying conjugation efficiencies. \textbf{(A)} The original cell locations \textbf{(left)}, and initial chemical concentration \textbf{(right)}. \textbf{(B-E)} On the \textbf{left}, the reconstructed cells, and on the \textbf{right}, the reconstructed chemical concentration, using \textbf{(B)} 100\% conjugation probability, \textbf{(C)} 40\% conjugation probability, \textbf{(D)}, 30\% conjugation probability, and \textbf{(E)} 20\% conjugation probability. The reconstructed chemical concentrations assumed the pioneer cells had 50 base pairs to encode the concentration. The black area on the outside is a border, not a chemical concentration. } \label{fig:S1}
\end{figure}

\begin{figure}[H]
\centering
\includegraphics[scale=.82]{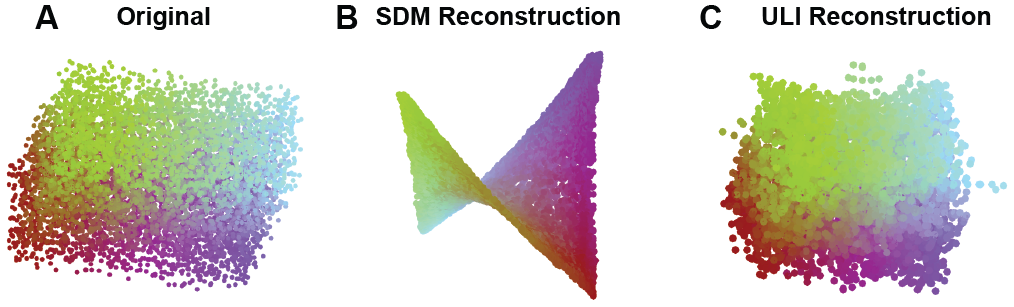}
\caption[Reconstruction of Non-cubes]{\textbf{Reconstruction of Non-cubes.} We do voxel puzzling as in \autoref{fig:3}A,B, with 8000 \SI{5}{\micro\meter} voxels. However, now the overall shape is a rectangular prism (height is half of the length and width) rather than a cube. \textbf{(A)} The original voxel locations. \textbf{(B)} Reconstruction using SDM. \textbf{(C)} Reconstruction using ULI.} \label{fig:S2}
\end{figure}

The simulations in the main text (\autoref{fig:7}) assumed that when an F\textsuperscript{+} and F\textsuperscript{-} cell are in contact, transfer of genetic information occurs 100\% of the time. Evidence to the contrary exists, especially in the absence of selection, where the fitness penalty associated with F\textsuperscript{+} strains carrying a plasmid, and the competition between cell growth and conjugation, limit plasmid transfer~\cite{ponciano07}. Furthermore, recombination of the F-plasmid with the host genome can result in Hfr (high frequency of recombination) strains, which have the ability to transfer genetic information (e.g. barcodes), but generally do not confer conjugative ability to the recipient~\cite{gupta09}. Our analysis, however, does not rely on every neighboring F\textsuperscript{+}/F\textsuperscript{-} pair to mate. Rather, it only requires that the descendants of nearby pioneer cells mate at least one once (if they're an F\textsuperscript{+}/F\textsuperscript{-} pair). In simulations, relatively faithful chemical reconstruction was accomplished with a conjugation probability of only 30\% (\autoref{fig:S1}).  This probability is likely very feasible, as neighboring pairs will have many chances to mate.

\subsection{Reconstruction of Non-cubes}

\subsection{Simulation Limitations}
\subsubsection{Neural Voxel Puzzling}
In our simulations, we aimed to demonstrate some of the problems that could arise when using real data. For instance, for voxel puzzling, we removed a large portion of the voxels to represent cell bodies taking up multiple voxels. It would likely be possible to place these voxels when using real data, by looking at the surrounding voxels that share a neuron tag.

In our voxel puzzling simulations, there were equal densities of neurons in all portions of the volume, which is not the case in the brain. This should not be a large concern, however, due to our method of constructing the similarity matrix. When we do thresholding, we are essentially taking the approximate nearest neighbors. This step removes the effects of having more total connections between voxels in one region (due to a high density of neurons) compared to another. 

Our neurons were also all oriented at random angles, so that voxels that were horizontal of each other had the same probability of sharing a neuron as voxels that were vertical of each other (or on top of each other). If this is not the case (e.g. axons are oriented in a similar manner in many brain regions), the reconstruction will be skewed. For instance, if most neurons were oriented vertically, in the reconstruction, the vertical component will be compressed. Following reconstruction, if it's seen that many neurons in a region are all oriented in the same manner, the dimensions of this region could be scaled proportionally. Moreover, having molecular annotations about whether axons or dendrites of a neuron are in a voxel could help with this process.

\subsubsection{Neural Connectomics Puzzling}
In our simulations of neural connectomics puzzling, we demonstrated that accurate reconstructions were still possible when there were multiple connection probability distributions (as a function of distance). However, this simulation only had 2 layers, and the connection probability distributions were consistent in each layer. It is possible that reconstructions could be less accurate when using more layers. This could be dealt with by using \textit{a priori} knowledge about neuroanatomy in order to ensure cubes were cut to contain a homogenous population. It also could be problematic if connection probability distributions differed within a layer across multiple cell types~\cite{liley94}. This could be dealt with by using annotated molecular information about cell types~\cite{usoskin14}. In this scenario, separate reconstructions could be done using neuronal populations with homogenous connectivity probability functions, and then these reconstructions could be combined. Additionally, it may be possible to estimate cell types given connectivity information (by using clustering methods on the connectivity data)~\cite{jonas14}, and then use these cell type estimates (as mentioned above) to help localize the neurons.

It is also important to note that connectomics puzzling is based on the idea that there is a higher probability of connection between neurons that are closer to each other. If this is not the case, then the connectomics puzzling reconstruction may be inaccurate. For certain cell types or portions of the brain (e.g. layer 5 pyramidal cells), there may be a higher probability of long-range connections than short connections. Using \textit{a priori} knowledge about neuroanatomy or information about cell types based on molecular annotations, these neurons could be excluded (or specific connections these neurons have could be excluded). Alternatively, this information about long-range connections could be utilized in connectomics puzzling to help provide an accurate reconstruction. 

\subsubsection{Chemical Puzzling}

In our simulations of chemical puzzling, we used simplified models of bacterial growth and movement. In terms of cell growth, we assumed that cell colonies keep growing outward and stop once they contact another colony. In reality, cells will push on one another, causing the colonies to move. Additionally, in our two dimensional simulations, we ignored the possibility of cells growing in the vertical dimension, which could lead to additional conjugations. While we could incorporate a more sophisticated (and much more expensive in terms of CPU time) simulation of bacterial growth, it is important to note that these simplifications will have little effect on whether two colonies have a conjugation, which is the metric that influences our results.

Our simulations also assumed that cells are stationary; that is, once a cell occupies a particular spot in two (or three) dimensional space, it stays there forever. In this case, the trajectory of a lineage is determined by a two (or three) dimensional random walk with uniform and unity step size, where two trajectories do not cross. This would certainly not be true in the case of motile cells or when there is sufficient mixing, as cells arising from different lineages could cross before dividing. Further, if the particular chemical being sensed induces a bias in cell motility (i.e. is a chemoattractant or chemorepellant), a bias will be further introduced into the colony growth model. These movement affects could potentially be problematic for two reasons: 1) it could allow far-away colonies to conjugate with one another; 2) pioneer cells could determine the chemical concentration of a location different from the location where they start to divide. Detailed analysis of these movement affects is beyond the scope of this paper, and we simply note that in the case that cell motility is also an unbiased random walk (a good assumption in the absence of chemotaxis and viscous flow~\cite{saragosti12}) and there is no mixing, our analysis stands in the limit that the average distance covered by the cells during one division time is much smaller than their size. Outside of this limit, our reconstructed images would be blurred.

\bibliographystyle{unsrt}
\bibliography{PuzzlingLib}

\end{document}